\documentstyle[aas2pp4,flushrt]{article}
\newcommand{\gw}{GW46.4+5.5}
\newcommand{\mum}{$\mu$m}
\newcommand{\ihund}{$I_{100}$}
\newcommand{\irexcess}{IR-excess}
\newcommand{\irexc}{IR~excess}
\newcommand{\iexcess}{\ihund-excess}
\newcommand{\iexc}{\ihund~excess}
\newcommand{\isixty}{$I_{60}$}
\newcommand{\thund}{$\tau_{100}$}
\newcommand{\texcess}{\thund-excess}
\newcommand{\texc}{\thund~excess}
\newcommand{\lb}{($l$, $b$)}
\newcommand{\degr}{$^\circ$}
\newcommand{\degrs}{$\!\!^\circ$}
\newcommand{\kms}{km~s$^{-1}$}
\newcommand{\vlsr}{$v_{\rm LSR}$}
\newcommand{\inhunit}{MJy~sr$^{-1}$~{(10$^{20}$~cm$^{-2}$)}$^{-1}$}
\newcommand{\cfunit}{cm$^{-2}$~{(K~\kms)}$^{-1}$}
\newcommand{\nh}{$N({\rm H})$}
\newcommand{\nhi}{$N({\rm HI})$}
\newcommand{\nhtwo}{$N({\rm H_2})$}

\newcommand{\lsol}{$L_\odot$}
\newcommand{\td}{$T_{\rm d}$}
\newcommand{\thi}{$T_{\rm HI}$}
\newcommand{\thtwo}{$T_{\rm H_2}$}

\newcommand{\tr}{$T^*_{\rm R}$}
\newcommand{\wco}{$W_{\rm CO}$}

\newcommand{\delvfwhm}{$\Delta v_{\rm FWHM}$}
\begin{document}

% title

\title{Infrared Excess and Molecular Gas in the Galactic Worm \gw}
\author{Kee-Tae Kim, Jeong-Eun Lee, and Bon-Chul Koo}
\affil{Department of Astronomy, Seoul National University, Seoul
151-742, Korea}

%\received{RECEIPT DATE}
%\accepted{ACCEPT DATE} 

\begin{abstract}

We have carried out high-resolution ($\sim$3$'$) HI and CO line
observations along one-dimensional cuts through 
the Galactic worm \gw. By comparing the HI data with $IRAS$ data,
we have derived the distributions of \iexc\ and \texc,
which are respectively the 100~\mum\ intensity and 100~\mum\ optical depth
in excess of what would be expected from HI emission.
In two observed regions, we were able to make a detailed
comparison of the infrared excess and the CO emission. We have found that
\texc\ has a very good correlation with the integrated intensity of CO
emission, \wco, but \iexc\ does not. There are two reasons for the poor
correlation between \iexc\ and \wco: firstly, there are regions with
enhanced infrared emissivity without CO, and secondly, dust grains
associated with molecular gas have a low infrared emissivity. In one 
region, these two factors completely hide the presence of molecular gas in
the infrared. In the second region, 
we could identify the area with molecular gas, but
\iexc\ significantly underestimates the column density of molecular
hydrogen because of the second factor mentioned above. 
We therefore conclude that 
\texc, rather than \iexc, is an accurate indicator of 
molecular content along the line of sight. 
We derive 
\thund/\nh=(1.00$\pm$0.02)$\times$10$^{-5}$~{(10$^{20}$~cm$^{-2}$)}$^{-1}$, 
and $X \equiv $\nhtwo/\wco$\simeq$0.7$\times$10$^{20}$ \cfunit. 
Our results suggest that \iexc\ could still be used to estimate the
molecular content if the result is multiplied by a correction factor
$\xi_{\rm c} \equiv
{<I_{100}/N{\rm (H)}>}_{\rm HI} / {<I_{100}/N{\rm (H)}>}_{\rm H_2}$
($\simeq 2$ in the second region), which accounts for the different infrared
emissivities of atomic and molecular gas. 
We also discuss some limitations of this work,
which could stem from using single-temperature model 
and the $IRAS$ 60~\mum\ intensity 
in estimating the dust optical depth along the line of sight.

\end{abstract}


\clearpage

%Text

% 1. Introduction
\section{Introduction}

In many studies of the interstellar medium,
it is essential to determine accurately the amount of molecular gas along
the line of sight.
However, the most abundant molecule, H$_2$, not only has a
large excitation temperature ($T_{\rm ex}\ge$509~K), which is not
usually attainable in a cold interstellar medium, 
but also has very small rotational
transition probabilities. In order to infer the H$_2$ column 
density \nhtwo, therefore, we generally rely on 
indirect methods. The most well-known indirect method for estimating the
H$_2$ column density is to observe other molecules such as CO, CS, or 
NH$_3$. 
%extinction measurements and other molecule observations. 
%If visual extinction were determined by star counts, the
%H$_2$ column density is derived using the dust-to gas ratio obtained
%empirically (Bohlin, Savage, $\&$ Drake 1978). This extinction method  
%H$_2$ column density is estimated from the column density of  
%Therefore other molecules instead have been observed 
%to infer the H$_2$ column density, \nhtwo. 
However, the fractional abundance of a molecule relative to H$_2$
is unlikely to be uniform from one cloud to another, 
or even sometimes within a cloud. 
In addition, there may be some differences among the
distributions of individual molecules because of their different photochemical
properties. In the case of CO, which has been widely used as a tracer of
H$_2$, the abundance varies from cloud to cloud
by up to three orders of magnitude depending on the visual extinction 
and astrochemical properties (Scoville $\&$ Sanders 1987;
van Dishoeck $\&$ Black 1988; Magnani $\&$ Onello 1995).

The infrared all-sky maps produced by the {\it InfraRed Astronomical
Satellite} ($IRAS$) mission presented a new opportunity to study 
the distribution of interstellar molecular gas. 
The Galactic radiation in the far-infrared (far-IR) appears to   
arise mostly from dust grains well-mixed with interstellar gas 
(Mathis, Mezger, $\&$ Panagia 1983). The $IRAS$ 100~\mum\ emission
intensity, \ihund, has been found to be tightly correlated with the total 
column density of interstellar gas at $|b| \geq$5\degr, and  
the 100~\mum\ emissivity per hydrogen nucleus, \ihund/\nh, seems to be
fairly uniform in the solar neighborhood : 
\ihund/\nh$\simeq$1~\inhunit~(de Vries et al. 1987; 
Boulanger $\&$ P\'{e}rault 1988; 
Heiles, Reach, $\&$ Koo 1988; Deul $\&$ Burton 1990, 1992).
Therefore, \iexc, i.e.,
the $IRAS$ 100~\mum\ emission in excess of what is expected 
from the HI column density, could be used as a tracer of H$_2$ in regions 
where the amount of ionized gas is negligible. 
Based on such an idea, several groups have identified
``infrared-excess (IR-excess)'' clouds in order to investigate the distribution
of molecular gas. However CO observational studies on
the \irexcess\ clouds showed that there is a substantial difference between 
the distribution of \irexcess\ clouds and that of CO-emitting clouds 
(D\'{e}sert, Bazell, $\&$ Boulanger 1988; 
Blitz, Bazell, $\&$ D\'{e}sert 1990; Heithausen et al. 1993; 
Reach, Koo, $\&$ Heiles 1994; Meyerdierks $\&$ Heithausen 1996; 
Reach, Wall, $\&$ Odegard 1998).
This discrepancy between the two distributions is in a sense
expected, because firstly the dust abundance and radiation-field
strength vary, and secondly the CO abundance is low
in diffuse molecular clouds, and finally the infrared emissivity of
the dust associated with molecular gas is low.
 
% t_100 - W_CO relation 

Another $IRAS$ method for tracing molecular gas is to use
100~\mum\ optical depth, \thund. This might be more
accurate than using 100~\mum\ intensity, at least in principle, because
optical depth is simply proportional to the amount of dust along the
line of sight if dust properties are the same. Indeed,
for several high-latitude clouds,
100~\mum\ (or 60~\mum) optical depth has been found to be correlated with
molecular gas column density better than the 100~\mum\ intensity alone
(Langer et al. 1989; Snell, Heyer, $\&$ Schloerb 1989;
Jarrett, Dickman, $\&$ Herbst 1989).
`IRAS' clouds similar to the \irexcess\ clouds have been identified
using this method, too (Wood et al. 1994).
These previous studies, which  were limited to high latitudes
($|b|$$\gtrsim$10\degr), however, simply
calculated dust optical depth after
subtracting flat background emission from the 60 and 100~\mum\
intensity
maps and showed that it had a very good correlation with
the integrated intensities of CO isotopes.
 
In this paper, we make a detailed comparison of 100~\mum\ intensity,
100~\mum\ optical depth, and the distribution of CO emission
along several one-dimensional cuts.
Our study differs from most previous studies in that we use both
high-resolution ($\sim$3$'$) HI and CO data in the analysis (cf. Reach
et al. 1994). The HI data make it possible to derive accurate
correlations between \thund\ (or \ihund) and \nhi, which is used to
identify `\texcess' (or \iexcess) regions and to estimate the excess
column densities of H nuclei. This excess H column density is compared
with the distribution of CO integrated intensity, \wco, to derive the
important conversion factor $X$$\equiv$\nhtwo/\wco. 
For comparison, most previous studies
(e.g., de Vries et al. 1987; Heithausen $\&$ Thaddeus 1990)
that derived the conversion factor from the $IRAS$ data have been
based on the comparison of \iexc\ and \wco, which may significantly
underestimate $X$ (Magnani $\&$ Onello 1995; Reach et al. 1998).
In this paper we compare these two approaches 
in detail and discuss their limitations.

The object that we study in this paper is the galactic worm \gw.
Galactic worms are wiggly, vertical structures which look like worms
crawling away from the Galactic plane in median-filtered HI maps
(Heiles 1984; Koo, Heiles, $\&$ Reach 1992).
\gw\ is a $\sim$8$^\circ$ long, filamentary structure
extending vertically from the Galactic plane in both far-IR (or HI) and
radio continuum emission (Figure 1). It is possible that \gw\ is 
the wall of a supershell
similar to the North Polar Spur, but at a greater distance
(Kim, Koo $\&$ Heiles 1999).
We describe the HI and CO line observations in Section 2.
In Section 3 we evaluate \ihund\ and \thund\ excesses
using the $IRAS$ and HI data, and compare their distributions
with that of \wco.
We discuss the nature of \iexc\ in \gw\ and some potential
problems of \texcess\ method in Section 4.
The main conclusions are summarized in Section 5.

% 2. Observations
\section{Observations}

HI 21~cm line observations were carried out 
using the 305 m telescope at 
Arecibo Observatory in 1990 October. The telescope had a HPBW of 3.$'$3
and a beam efficiency of 0.8 at 1.4 GHz (Reach, Koo, $\&$ Heiles 1994). 
We observed both circular
polarizations simultaneously using two 1024 channel correlators with 5
MHz bandwidth each, so that the velocity resolution was 2.03~\kms\ after
Hanning smoothing. Each spectrum was obtained by integrating for 1 minute
using frequency switching. We made a total of 9 one-dimensional cuts
through \gw\ at a set of constant galactic latitudes.
% given by 1\degr, 2\degr, $\cdots$, 9\degr.
The beam separation was 3$'$.
The positions of our one-dimensional cuts are listed in Table 1.
 
CO J=1$-$0 line observations were made in 1994 February
using the 4~m telescope (HPBW=2.$'$5) at Nagoya University in Japan.
We obtained 8 one-dimensional cuts in the same way as in the HI line
observations. The observed positions together with their reference positions
are summarized in Table 1.
An SIS mixer receiver and a 1664 channel Acousto-Optical Spectrometer (AOS)
with 40 MHz bandwidth were used. The velocity resolution was
0.67~\kms\ after Gaussian smoothing. The system temperature
varied in the range 480$-$700~K during the observing sessions
depending on weather conditions and elevation of the source.
Absolute-position switching
instead of frequency switching was used in order to
prevent the contamination of spectra
by the atmospheric CO emission. The reference positions were checked to
be free of appreciable (\tr$<$0.1~K) CO emission. The velocity was
centered at \vlsr=40~\kms\ and the velocity coverage was from $-$19 to
+91~\kms. The on-source integration time
was 2 minutes and the typical rms noise level was 0.2~K per channel after
Gaussian smoothing.
The intensity scale was calibrated with respect to the standard
source S140 which was assumed to have \tr=20~K (Yang $\&$ Fukui
1992).

% 3. Results
\section{Infrared Excess and Molecular Gas}

\subsection{HI Gas and Infrared Excess}
 
Figure 2 shows a sample from our HI spectrum 
at \lb=(45.$^\circ$10, 4.$^\circ$00)
where the strongest CO line emission was detected.
There are three peaks at positive velocities in the HI spectrum 
and the peaks in the velocity range \vlsr$\simeq$18$-$40~\kms\ are the
components associated with \gw\ (Kim et al. 1999). 
The components at negative velocities might be from the warped 
Galactic plane outside of the solar circle. 
The physical and dynamical properties of the worm will be discussed 
in a separate paper (Kim et al. 1999). 

We assume that the HI emission is optically thin and compute 
the HI column density along a given line of sight, $N$(HI), from
 
\begin{equation}
N({\rm HI})~=~1.822~\times~10^{18}~\int{T_{\rm b}}~{\rm d}v~~~~{\rm cm}^{-2},
\end{equation}
 
\noindent
where $T_{\rm b}$ is the brightness temperature in K and $v$ is 
the velocity in \kms. Our assumption seems to be valid because 
the peak temperature of the HI line is much lower 
than the typical spin temperature of HI gas,
$T_{\rm s}$=125~K, for most sight lines. The integral range is from 
% over the velocity range, 
\vlsr=$-$100 to +150~\kms. 
% The atomic gas content was known to be proportional to integrated HI
% line intensity over the Galaxy except some regions in the Galactic plane,
% i.e., 10\degr$<|l|<$70\degr\ at $|b|<$0.\degrs5 and
% $l$=0\degr$\pm$10\degr, $l$=180\degr$\pm$10\degr,
% and 70\degr$<|l|<$80\degr\ at $|b|<$3\degr\ (Deul $\&$ Burton 1992).
% (justifications for assumption of optically thin)
Figure 3a is the plot of the 100 \mum\ intensity, \ihund, against  $N$(HI) in
the observed regions.
We can see a strong correlation between the two physical parameters. 
Different beam sizes of the HI and $IRAS$ observations could introduce some
scattered data points in the figure. But the effect must be small
because the beam sizes are roughly the same ($\sim$3$'$).
The figure shows only the data from the regions at $b \ge$3\degr\
where both HI and CO line observations were made. 
The data at lower galactic latitudes are
not used because of the possible contribution from dust associated with
ionized gas. The open circles, squares, and pentagons represent the
points with detectable CO emission at $b$~=~3\degr, 4\degr, and 5\degr,
respectively. The area of each symbol is proportional to the
integrated CO intensity. The crosses represent the points without
detectable CO.
% implication of y-intercept
A least-squares fit excluding the points with detectable CO yields 
(\ihund/MJy~sr$^{-1}$)~=~(1.32$\pm$0.02)
(\nhi/10$^{20}$~cm$^{-2}$)~$-$~3.37$\pm$0.7.
Our fitting procedure involves two steps: 
(1) All the crosses are fitted to a straight line, and 
(2) the crosses within $\pm$2$\sigma$ deviation from the first fit
are fitted to a new straight line.
% We performed the least-squares fit in two steps: 
% (1) We fit to all the crosses; (2) We fit to only the crosses within
% $\pm$2$\sigma$ deviation from the first fit.
The negative zero-intercept could be from errors in the subtraction of 
zodiacal light.
The derived \ihund/$N$(HI) agrees with the results of 
previous observational studies. For example, Boulanger et al. (1986), 
de Vries et al (1987), and Heiles et al. (1988) obtained 
1.4$\pm$0.3, 1.0$\pm$0.4, and 
$\sim$1.3 MJy~sr$^{-1}$~{(10$^{20}$~cm$^{-2}$)}$^{-1}$, respectively.

We now compare the HI column density and 100~\mum\ optical depth. 
Assuming that the far-IR emission is
optically thin and that the temperature of dust grains is constant along
the line of sight, the 60/100 \mum\ color temperature, \td, can be derived
according to the formula
 
\begin{equation}
T_{\rm d} = {c_1 \biggl[\frac{1}{\lambda_{\rm 60}} -
\frac{1}{\lambda_{\rm 100}} \biggr]}/
{{\rm ln} \biggl[\frac{I_{\rm 100}}{I_{\rm 60}}
{\biggl(\frac{\lambda_{\rm 100}}{\lambda_{\rm 60}} \biggr)}^{n+3}
\biggr]},
\end{equation}
 
\noindent
where $c_1$=$hc/k_{\rm B}$=1.441
%, $I_{\lambda}$ is the intensity at a given wavelength, 
and $n$ is the index in
the emissivity law, $Q_{\rm abs}(\lambda) \sim \lambda^{-n}$.
For the graphite-silicate dust grain model of Mathis, Rumpl,
$\&$ Norsdsieck (1977; hereafter MRN dust model), $n$ lies between 
1 and 2 (Draine $\&$ Lee 1984). We take $n$=1.5 in this paper.
The 100 \mum\ optical depth, \thund, is derived from 
\thund=\ihund/$B_{\rm \lambda}$(\td), where $B_{\rm \lambda}$(T) 
is the Planck function.
Figure 3b displays the 100 \mum\ optical depth versus 
the HI column density. A strong correlation also exists between these
two quantities.
A least-squares fit has been performed in a similar way to yield
(\thund/10$^{-5}$) = (1.00$\pm$0.02) (\nhi/10$^{20}$ cm$^{-2}$) + 0.01$\pm$0.6. 
The estimated \thund/\nh\ ratio is much smaller than the value 
$<\tau_{\rm 100}/N({\rm H})>$ = 6.3 $\times10^{-5}$ {(10$^{20}$ 
cm$^{-2}$)}$^{-1}$
obtained by Boulanger et al. (1996) 
from an analysis of the {\it Cosmic Background Explorer} ($COBE$) mission
% the Diffuse Infrared Background Experiment (DIRBE), 
% the Far-Infrared Absolute Spectrometer (FIRAS),
and the Leiden-Dwingeloo HI survey data.
This difference between the two values cannot be entirely due to the calibration
difference between the $IRAS$ and $COBE$ observations. 
Instead, the difference may be attributed to the effects of 
small, transiently heated dust grains (Langer et al. 1989)
and to the insensitivity of the $IRAS$ observations
to cold (\td$ <$15~K) dust grains
(Snell et al. 1989; Jarrett et al. 1989; Wood et al.  1994).

If the dust-to-gas ratio and the
infrared emissivity per hydrogen nucleus are uniform over the 
atomic and molecular regions, the infrared excess can be
used to estimate the H$_{\rm 2}$ column density. 
The infrared excess may be determined either from \ihund~or \thund\ using the
following formulae:
 
\begin{equation}
N({\rm H_{\rm 2}})_{I_{100}}~\equiv~\frac{1}{2}
\bigg[ \frac{I_{\rm 100,c}}{<I_{\rm 100,c}/N({\rm H})>}~-~N({\rm HI})
\biggr] ~~~~{\rm cm}^{-2} 
\end{equation}

\noindent
and

\begin{equation}
N({\rm H_{\rm 2}})_{\tau_{100}}~\equiv~\frac{1}{2}
\biggl[ \frac{\tau_{\rm 100,c}}{<\tau_{\rm 100,c}/N({\rm H})>}~-~N({\rm HI})
\biggr] ~~~~{\rm cm}^{-2},
\end{equation}

\noindent
where the subscript `c' indicates the quantity corrected for offset and
the angle bracket indicates an average ratio. Since we excluded the data
points with detectable CO, we may use the derived ratios as the ratios 
with respect to the column density of {\it hydrogen nuclei}, i.e.,
$<I_{\rm 100}/N({\rm H})>$=1.32
MJy~sr$^{-1}$~{(10$^{20}$~cm$^{-2}$)}$^{-1}$ 
and $<\tau_{\rm 100}/N({\rm H})>$=1.00$\times 10^{-5}$
{(10$^{20}$~cm$^{-2}$)}$^{-1}$, respectively.

\subsection{Comparison of Infrared Excess and CO Line Intensity}

We have detected CO emission in five regions (Table 2). 
Figure 2 shows the spectrum of the strongest CO line and compares it with
the corresponding HI spectrum. Note that the velocity range of our CO
observations was from $-$19 to +91~\kms, so that we could not detect the
CO emission, if any, associated with the HI gas at negative velocities.
It is, however, very unlikely that there is appreciable CO emission in
this outer part of the Galaxy. We, in fact, made some test observations
with enough velocity coverage, but could not detect any emission
(\tr$<$0.2~K). In any case, we believe that we have detected the
emission from most of the CO gas.
The CO emission from regions C and D is almost certainly
associated with \gw, while that from regions A and B is probably not
because the velocities are very different.
The CO emission from region E is also possibly related to
the worm although its central velocity is somewhat lower than the
velocity of the worm. The physical association of molecular gas
with \gw\ will be discussed in a separate paper (Kim et al. 1999).
As we have mentioned in Section 3.1, the regions at $b$=1\degr\ (A and B)
are not included in our analysis because of the complication due to the
presence of ionized gas. Region E is not included because it has
only a few observed points. We thus limit our subsequent 
analysis to regions C and D.
 
Figures 4a and 4b compare the values of $N$(H$_{\rm 2}$) 
derived from equations (3)
and (4) with the integrated CO line intensity 
in regions C and D, respectively. The integral ranges are from \vlsr=+20
to +40~\kms\ for region C and from \vlsr=+15 to +30~\kms\ for region D.
These are velocity ranges where the CO emission is detected.
In region C, the \nhtwo$_{\tau_{100}}$ distribution matches very 
well that of \wco, but the
\nhtwo$_{I_{100}}$ distribution does {\it not}. The \nhtwo$_{I_{100}}$ 
distribution differs from that of \wco\ significantly 
at $b$$\simeq$46\degr$-$47\degr. 
In region D, on the other hand, both \nhtwo$_{\tau_{100}}$ 
and \nhtwo$_{I_{100}}$ 
match very well with \wco\ although their absolute scales differ by 
about a factor of 2. Hence, \texc\ appears to trace molecular 
gas better than \iexc, which is also obvious from Figures 3a and 3b 
since the points with CO are separated more clearly from those 
without CO in the latter.
This is more clearly shown in Figure 5, where \nhtwo$_{\tau_{100}}$ and
\nhtwo$_{I_{100}}$ are compared with \wco. \nhtwo$_{\tau_{100}}$ is
obviously proportional to \wco\ whereas \nhtwo$_{I_{100}}$ has 
much weaker correlation with \wco. 
We determined 
\nhtwo$_{\tau_{100}}$/\wco=(0.70$\pm$0.05)$\times 10^{20}$ \cfunit\  
by a least-squares fit. 
The error quoted represents only a statistical error in the fit.
% This value may be an upper limit for \nhtwo/\wco\ because, as mentioned in
% Section 2, the HI beam is a little larger than the CO one.
This conversion factor is much smaller than the
estimated value for molecular clouds in the Galactic plane,
$X$=(1.8$-$4.8) $\times$ 10$^{20}$ \cfunit\ (Scoville $\&$ Sanders 1987),
but comparable to that of high latitude clouds,
% consistent with the results on high latitude clouds,
$X$ $\simeq$ 0.5 $\times$ 10$^{20}$ \cfunit\ 
(e.g., de Vries et al. 1987; Heithausen $\&$ Thaddeus 1990; Reach et al. 1998). 
% The estimated value may be an upper limit for \nhtwo/\wco\ because, 
% as mentioned in Section 2, the HI beam is a little larger than the CO one.
Since there seems to be a systematic difference between the correlations
in regions C and D,
we also determined the slopes of the correlations separately.
They are (0.40$\pm$0.05) and (0.93$\pm$0.06)$\times 10^{20}$ \cfunit,
respectively. This difference could be due to variations in the
dust-to-gas ratio (Magnani $\&$ Onello 1995) or due to variations of
the gas-phase carbon abundance (Heithausen $\&$ Mebold 1989). 
Further study is needed to address this question.

As we will discuss in more detail
in Section 4.2, the correlation between \wco\ and
\nhtwo$_{I_{100}}$ is not apparent partly 
because of the existence of a local heating source in region C.
If we limit the analysis 
to region D, there is a weak correlation between the two quantities,
\nhtwo$_{I_{100}}$/\wco = (0.59$\pm$0.05) $\times 10^{20}$ \cfunit. 
Note that the
coefficient of proportionality is about 1/2 that of the
\wco-\nhtwo$_{\tau_{100}}$ relationship, the significance of which will be
discussed in next section.

% 4. Discussion
\section{Discussions}

\subsection{\irexc\ as a Molecular Tracer}

\noindent
\centerline{\it 4.1.1~~\iexc\ as a molecular tracer}
% \subsubsection{\iexc\ as a molecular tracer}

We have found that the \iexcess\ distribution does not correlate well with
the CO distribution. The discrepancy between the two distributions has
also been noted in several previous studies (D\'{e}sert et al.
1988; Blitz et al. 1990; Heithausen et al. 1993; Reach et al. 1994; 
Meyerdierks $\&$ Heithausen 1996; Reach et al. 1998). 
In general, the regions with such discrepancies are divided
into two categories: (a) \iexcess\ regions without CO emission, and (b)
CO-emitting regions without \iexc. The sources that fall into the first
category could be regions with enhanced 100~\mum\ emissivity.
The 100~\mum\ emissivity is generally dependent on the
dust-to-gas ratio and dust temperature.
It has been found that the dust-to-gas ratio changes with position on
an angular scale of $\sim$10\degr\ and by up to a factor of 4
(Burstein $\&$ Heiles 1978; D\'{e}sert et al. 1988).
Variation in the dust-to-gas ratios on smaller scales is expected, too.
The steady-state temperature of dust grains is determined by
their optical properties and the intensity of radiation field (e.g.,
Draine
$\&$ Anderson 1985). The \iexcess\ region at $l$$\simeq$46\degr\ in
region C is due to the strong radiation field from a nearby O star 
(see Section 4.2).
Alternatively, the sources could be diffuse molecular
clouds without detectable CO (Blitz et al.
1990; Reach et al. 1994; Meyerdierks $\&$ Heithausen 1996;
Reach et al. 1998).
This is possible
because CO may not be shielded by HI, H$_2$, and
dust from photodissociation, and is unlikely to be collisionally excited
in the low-density region.
On the other hand,
the sources in the second category, i.e., molecular clouds without
\iexc, could be present because of the low emissivity of dust grains
associated with molecular gas.
If there were no embedded heating source in the molecular cloud,
the dust temperature drops from the surface to the center of the cloud.
According to Mathis et al.  (1983), for example, the dust
temperature drops by about 3~K from the edge to the center for a cloud
with $A_{\rm v}$=5~mag.
The 100~\mum\ emissivity of dust grains is very sensitive to
dust temperature and changes by about 40\%
as dust temperature changes by 1~K over the range
18$\leq T_{\rm d} \leq$25~K. At $l$$\simeq$46.\degrs3 in region C, the
low emissivity almost completely counterbalances the significant
fraction of \iexc\ from dust grains associated with molecular gas.

The low far-IR emissivity of molecular clouds has another important
implication for using \iexc\ as a molecular tracer. {\it The
\iexc\ significantly underestimates the H$_2$ column density.} In region D,
the H$_2$ column density derived from \iexc\ is smaller than that derived
from \texc\ by about a factor of 2. This is because in equation (3) we
used $<I_{100}/N{\rm (H)}>$ obtained from 
HI line observations. If there is a
molecular cloud along the line of sight, a correct expression for the
observed 100~\mum\ emission would be

\begin{equation}
I_{100} = {<I_{100}/N{\rm (H)}>}_{\rm HI} N{\rm (HI)}
+ 2{<I_{100}/N{\rm (H)}>}_{\rm H_2} N{\rm (H_2)},
\end{equation}

\noindent
where ${<I_{100}/N{\rm (H)}>}_{\rm HI}$ and 
${<I_{100}/N{\rm (H)}>}_{\rm H_2}$ are, respectively,
100~\mum\ emissivities per hydrogen nucleus in atomic and molecular regions. 
Hence, we need to apply a correction factor $\xi_{\rm c}$ 
to \nhtwo$_{I_{100}}$ in equation (3) in order to obtain 
an accurate H$_2$ column density: 

\begin{equation}
\xi_{\rm c} \equiv 
{<I_{100}/N{\rm (H)}>}_{\rm HI} / {<I_{100}/N{\rm (H)}>}_{\rm H_2}.
\end{equation}

\noindent
Since ${<I_{100}/N{\rm (H)}>}_{\rm HI}$ is in general greater than 
${<I_{100}/N{\rm (H)}>}_{\rm H_2}$, $\xi_{\rm c} \ge$1 and, in \gw,
$\xi_{\rm c} \simeq$2. 
Our correction factor $\xi_{\rm c}$ is equivalent to 
${<B_\nu (T_{\rm d})>}_{\rm HI}$ /${<B_\nu (T_{\rm d})>}_{\rm H_2}$
%where $B_\nu (T_{\rm d})$ is the Planck function, 
used by Magnani $\&$
Onello (1995) and Reach et al. (1998). From their study of \iexcess\ clouds,
Reach et al. (1998), for example, found $\xi_{\rm c}$=3.8 as an average
value for the solar neighborhood.  
For the inner Galaxy, if we use the mean
temperature of ${<T_{\rm d}>}_{\rm HI}$=21.0$\pm$1.0~K and 
${<T_{\rm d}>}_{\rm H_2}$=19.0$\pm$1.0~K derived 
by Sodroski et al. (1994) using 
the Diffuse Infrared Background Experiment (DIRBE) 
140 and 240~\mum\ observations, 
we obtain $\xi_{\rm c}$=2.1. Our result for $\xi_{\rm c}$, however,
cannot be compared directly with these results because our study is
based on the $IRAS$ 60~\mum\ brightness which is largely affected by the
emission from small grains.
Since $B_\nu (T_{\rm d})$ is very sensitive
to \td, it is necessary to estimate the correction factor
independently in studying specific regions (Lee, Kim, $\&$ Koo 1999).

\noindent
\centerline{\it 4.1.2~~\texc\ as a molecular tracer}
% \subsubsection{\texc\ as a molecular tracer}

According to our results, \texc\ appears to be an accurate indicator of
the molecular content along the line of sight; it not only traces the
presence of molecular gas correctly but also increases linearly with the
amount of molecular gas. There are two reasons for this. 
First it separates the
intensity peaks produced by enhanced infrared emissivity. Second it
corrects for the low infrared emissivity of molecular clouds. 
In the present work, however, there are two limitations as discussed
below.

The first limitation stems from using a mean temperature in deriving
\thund\ along the line of sight.
The expression for \nhtwo\ in equation (4) is, in fact, an exact one if dust
properties and the dust-to-gas ratio remain the same in atomic and
molecular regions. In practice, however, there is an error because, in
deriving \thund\ on the right hand side of equation (4), we have used an
average color temperature along the line of sight (defined in equation (2))
even if we need to use different color temperatures 
in deriving the optical depths of atomic and molecular regions. 
% We now estimate the error in total optical depth produced by using a
% mean color temperature for a simple case.
It is straightforward to estimate the error when there are only two dust
components with different emissivities along the line of sight.
If there is atomic gas of optical depth \thund(HI)
at color temperature \thi\ and molecular gas of optical depth
\thund(H$_2$) at \thtwo($<$\thi), the ratio of the optical depth $\tau_{100}'$,
derived by using a mean color temperature, to the true optical depth 
$\tau_{100}$ is given by

\begin{equation}
% \frac{\tau_{100}'}{[\tau_{100}({\rm HI})+\tau_{100}({\rm H_2})}
\frac{\tau_{100}'}{\tau_{100}}
= { \biggl [ \frac{1+r~{\rm exp}(-\triangle t)}
{1+r~{\rm exp}(-\frac{\triangle t}{\lambda_{60}/\lambda_{100}})}
\biggr ] }^{\frac{\lambda_{60}/\lambda_{100}}
{1 - \lambda_{60}/\lambda_{100}}}
\frac{1+r~{\rm exp}(-\triangle t)}{1+r},
\end{equation}

\noindent
where $r \equiv \tau_{100}({\rm H_2})/\tau_{100}({\rm HI})$,
$\triangle t \equiv$ 
($c_1/\lambda_{\rm 100}$)(1/${T_{\rm H_2}}$ $-$ 1/${T_{\rm HI}}$),
and the other parameters have the same meaning as in Section 3.
The derived optical depth is always less than the true value  
and, as expected, 
the difference converges to zero as $r$ approaches zero or infinity.
% $\tau_{100}' / \tau_{100}$ is always less than 1 and,
% as expected, converges to 1 as $r$ approaches zero or infinity.
Figure 6 shows, as solid lines, how $\tau_{100}' / \tau_{100}$ 
varies in the ($r$, $\triangle t$) plane.
The ratio $\tau_{100}' / \tau_{100}$ is affected more strongly by the warmer 
dust associated with atomic gas in the sense that the error in optical depth 
is larger at $\tau_{100}({\rm H_2})/\tau_{100}({\rm HI})$=$r_{\rm o}$
($>$1) than that at $\tau_{100}({\rm H_2})/\tau_{100}({\rm HI})$=1/$r_{\rm o}$
for a given value of $\triangle t$. Similar figures can be found in 
Langer et al. (1989) and Snell et al. (1989), but Figure 6 is a
generalized one. 
The error in optical depth would induce an error in \nhtwo$_{\tau_{100}}$. 
The error can be determined by the following formula:

\begin{equation}
\frac{N({\rm H_2})_{\tau_{100}'}}{N({\rm H_2})_{\tau_{100}}}
= \frac{\tau_{100}'/\tau_{100}-1/(1+r)}
{r/(1+r)}.
\end{equation}  

\noindent
The dependence of \nhtwo$_{\tau_{100}'}$/\nhtwo$_{\tau_{100}}$ on 
($r$, $\triangle t$) is displayed by dotted lines in Figure 6. 
% The error in \nhtwo$_{\tau_{100}}$ is in general larger than 
% that in $\tau_{100}$ for a given pair of ($r$, $\triangle t$).
The ratio \nhtwo$_{\tau_{100}'}$/\nhtwo$_{\tau_{100}}$ 
is in general smaller than 
$\tau_{100}' / \tau_{100}$ for a given ($r$, $\triangle t$) pair.
The most striking difference between the two quantities is that, 
as $r$ approaches zero, the error in \nhtwo$_{\tau_{100}}$
remains constant for given $\triangle t$ whereas that in $\tau_{100}$ 
tends to zero.
% As $r$ approaches zero, moreover, 
% \nhtwo$_{\tau_{100}'}$/\nhtwo$_{\tau_{100}}$
% remains constant ($<$1) for a given 
% $\triangle t$ while $\tau_{100}' / \tau_{100}$ increases to 1.
Since the visual extinction is $A_V$$\simeq$1.5~mag at the CO peak in
region D using \nh/$A_V$=1.9$\times$10$^{21}$~cm$^{-21}$~mag$^{-1}$ 
(Bohlin, Savage, $\&$ Drake 1978), \thtwo\ is expected to be 
$<$3~K lower than \thi\ (Mathis et al. 1983). 
Therefore, 
the error in \nhtwo$_{\tau_{100}}$ may be $\lesssim$20\%
% \nhtwo$_{\tau_{100}'}$/\nhtwo$_{\tau_{100}} \gtrsim$0.8 
in region D where
$\tau_{100}({\rm H_2})/\tau_{100}({\rm HI}) \simeq$1/2 (See Figure 6).
The error in the correction factor $\xi_{\rm c}$ is directly related to that
in \nhtwo$_{\tau_{100}}$ because it is actually derived
from \nhtwo$_{\tau_{100}'}$/\nhtwo$_{I_{100}}$: 
$\xi_{\rm c}' / \xi_{\rm c}$=\nhtwo$_{\tau_{100}'}$/\nhtwo$_{\tau_{100}}$.

The second limitation derives from using \isixty\ in estimating the dust
optical depth. Interstellar dust grains are known to be composed of
small, transiently heated grains and classical, large grains in thermal
equilibrium
(Draine $\&$ Anderson 1985; 
D\'{e}sert, Boulanger, $\&$ Puget 1990). 
The emission at 60~\mum\ comes from both small grains
and large grains, while the emission at
100~\mum\ originates mainly from large grains.
D\'{e}sert et al. (1990) divided  dust grains into
three components: Polycyclic Aromatic Hydrocarbins (PAHs), Very Small
Grains (VSGs), and Big Grains (BGs). 
In their model,
the contributions of VSGs are $\sim$60\% at 60~\mum\ and
$\sim$15\% at 100~\mum\ for the solar neighborhood.
Sodroski et al. (1994) found from an analysis of the DIRBE 
and IRAS data that the average values of the contribution
of small grains at 60~\mum\ are $\sim$40\% and $\sim$60\% in the inner
and outer Galaxy, respectively. 
% Hence, about half of the
% 60~\mum\ emission seems to be attributed to small grains.
$IRAS$ studies of nearby molecular clouds suggest that the small grains
are present largely in the halo surrounding the clouds and that their
relative abundances vary (Boulanger et al. 1990; 
Laureijs, Clark, $\&$ Prusti 1991; Bernard, Boulanger,
Puget 1993). Therefore, \thund\ that we have derived using
\isixty/\ihund\ does {\it not} represent 
the usual ``optical depth" of
dust grains. Instead it might be largely affected by small grains with
varying abundances. However, the fact that we do observe a very good
correlation between \texc\ and \wco\ (or \thund\ and \nhi)
seems to indicate 
that \thund\ is still proportional to the amount of dust along the line
of sight.
Compared to \ihund,
\thund\ can be interpreted as a quantity corrected for variation in
the abundance of small grains as well as variation in the
temperature of large grains.
%that the abundance
%of small grains is correlated with H$_2$ column density, i.e., their
%abundance is lower when the H$_2$ column density is greater. This
%implication appears to be consistent with the results of other studies
%(Boulanger et al. 1990; Laureijs, Clark, $\&$ Prusti 1991).
If one uses far-IR data at wavelengths longer than 100~\mum,
such as the DIRBE data for the 100, 140, and 240~\mum\ wavebands, one can
resolve this problem. In the present days, however, 
the IRAS data have the highest angular resolution 
among the available infrared survey data.

In the past decade, several groups have found for high-latitude
clouds that \thund\ (or $\tau_{60}$) is better correlated 
with the gas column density than \ihund\
(Langer et al. 1989; Snell et al. 1989; Jarrett et al. 1989; Wood et al.
1994). At high latitudes, however,
it is unlikely that the \texcess\ distribution is
markedly different from the \iexcess\ distribution, because
most high-latitude clouds are diffuse and quiescent compared to
the molecular clouds near the Galactic plane.
In addition, the low signal-to-noise ratio of \isixty\ could introduce a
large error in \thund. At low latitudes, on the contrary, 
the \texcess\ distribution is expected to
represent the distribution of molecular gas much better than the
\iexcess\ distribution, because
the number density of luminous stars increases and most molecular clouds
contain high-extinction regions and/or embedded luminous stars.

\subsection{Local Heating Source in \gw}

In the case of region C, the \iexcess\ distribution is largely affected by
the variation of dust temperature (Figure 4a).  The \iexcess\ peak at
$b$=46.\degrs0 appears not to be due to dust grains associated
with molecular gas but to an increase in dust
temperature. We found a massive star, LSII +12.\degrs3, 
near the position of interest, \lb~=~(46.\degrs0, 3.\degrs0) (Drilling 1975). 
The star was found to be a luminous $blue$ giant star,
O9~III (Vijapurkar $\&$ Drilling 1993). The coordinates of the star are 
($\alpha$, $\beta$)$_{1950}$=(19:03:14.3, 12:46:21) or 
\lb=(45.\degrs97, 2.\degrs75).
An infrared point source (IRAS19031+1247) and a radio point source 
with a diffuse extended envelope (F$\ddot{\rm u}$rst et al. 1990) lie in
the vicinity of the star.
IRAS19031+1247 seems to be an OH/IR star on the basis of its position in
the (60$-$25)~\mum\ versus (25$-$12)~\mum\ 
color-color diagram (cf. Lewis 1994). 
In order to reveal the physical relationship between the two
sources and the star, further studies are required.
We estimate the distance of the massive star to be 
$d$=2.2$\pm$0.8~kpc, using the spectroscopic data mentioned above and
the $UBV$ photometric data of Drilling (1975), $V$=10.$\!\!^m$72,
$B-V$=1.$\!\!^m$00. Here we adopt 
$M_V$=$-$5.$\!\!^m$3$\pm$0.$\!\!^m$7 for O9~III
(Conti et al. 1983) and $R \equiv A_V$/$E_{B-V}$=3.3 because $R$ for the
region is likely to be somewhat higher than the average value for 
the Galactic plane (Turner 1976). 
The estimated value is barely in agreement with the kinematic
distance of HI gas associated with \gw, $d_{\rm kin}$=1.4~kpc (Kim et
al. 1999).
 
Figure 7 displays the large-scale distribution of 
the \isixty/\ihund~ratio around
region C. The \isixty/\ihund~ratio is enhanced around 
the massive star, which is 
% marked as a star symbol, 
located where two arrows meet, and decreases 
monotonically from the star to the position of interest.
Assuming that $Q_{\rm abs}(\lambda)$ varies as $\lambda^{-n}$,
the total excess infrared luminosity from
the \isixty/\ihund~ enhanced region can be derived from the following
formula:

\begin{equation}
L_{\rm IR}^{\rm ex}~=~4 \sigma d^2 \int_{\Omega_{\rm s}}
\Bigl[ <\tau>_{\rm T_{\rm d}} T_{\rm d}^4 - 
<\tau>_{\rm T{\rm d,B}} T_{\rm d,B}^4 \Bigr] d\Omega,
\end{equation}

\noindent
where $\sigma$ is the Stephen-Boltzmann constant, $d$ is the distance to 
the massive star, and the subscript B indicates quantities in the
absence of the star. The Planck-averaged optical depth is defined as
$<\tau>_{\rm T}$=${ \int \tau_{\lambda} B_{\lambda}({\rm T}) d\lambda}
/ {\int B_{\lambda}({\rm T}) d\lambda}$ ($\propto$~\td$^{1.5}$ for $n$=1.5).
Using $d$=2.2~kpc and $T_{\rm d,B}$=25.2~K (\isixty/\ihund=0.22 
with $n$=1.5),
% we estimated  total excess infrared luminosity from
% the \isixty/\ihund~ enhanced region. 
% The mean dust temperature was
% determined using mean excess intensities at 60 and 100~\mum. 
the total excess infrared
luminosity is $L_{\rm IR}^{\rm ex}$=6500~\lsol\ over a solid angle of
$\Omega_{\rm s}$=2.4$\times$10$^{-4}$~sr. This value is
an order of magnitude smaller than the stellar luminosity, 
$L_*$=2.2$\times$10$^5$~\lsol\ (Panagia 1973). 
% Further, it may be an upper limit to the total excess infrared 
% luminosity because the observed \isixty/\ihund~ratio enhancement 
% might be in part due to the enhancement of the abundance of small grains. 
Hence, the star has enough luminosity to produce the observed infrared
luminosity, and the \iexcess\ peak at $b$=46.\degrs0 in region C is probably 
due to heating by an O-type giant star.
%the dust grains in the case are heated by this star.

% 5. Conclusions
\section{Conclusions}

Infrared (IR) excess, defined as the emission in excess of what would be
expected from HI emission, has been used for estimating the amount of
molecular gas along the line of sight. However, at the same time, there have
been molecular, particularly CO, line studies showing that the method
does not necessarily yield reliable results. That is, there are both
\irexcess\ regions without detectable CO and CO-emitting regions without IR
excess. Region C in our study is a good example. The sources that fall
into the first category could be either regions with enhanced IR
emissivity or diffuse molecular clouds without CO emission. In region C,
it is excess heating from a nearby O star that produces an IR-excess
region. On the other hand, the sources that fall into the second
category could be molecular clouds with low IR emissivity. In region C,
the low emissivity completely hides the presence of a molecular cloud in
the infrared.

Our result shows that we can avoid the above confusion problem by
using the optical-depth excess instead of IR excess, i.e., the optical
depth in excess of what would be expected from HI emission. We have
found a very good correlation between 100~\mum\ optical-depth (\thund)
excess and integrated intensity of CO emission, \wco. We derive the
conversion factor between \nhtwo\ and \wco,
$X$$\simeq$0.7$\times$10$^{20}$ for the galactic worm \gw.
According to our result, however, the conversion factors in regions C
and D differ by about a factor of 2, although both regions are parts of
the worm. A more complete observation is necessary for the study of the
variation of the conversion factor.

Another merit of using optical-depth excess, which may be more
important, is that it gives 
an {\it accurate} value for the amount of molecular gas.
IR excess would significantly underestimate \nhtwo\ if one uses an IR
emissivity estimated from HI emission, 
${<I_{100}/N{\rm (H)}>}_{\rm HI}$. We have introduced a correction
factor 
$\xi_{\rm c} \equiv
{<I_{100}/N{\rm (H)}>}_{\rm HI} / {<I_{100}/N{\rm (H)}>}_{\rm H_2}$,
which should be applied in order to account for the different IR
emissivity of molecular clouds. In general $\xi_c \ge 1$ because dust
grains in molecular clouds are usually colder than dust grains in atomic
gas. In region D in our study, $\xi_c \simeq 2$. We should therefore be
rather cautious in converting IR excess to \nhtwo.

Two potential problems in using optical-depth excess, however, would be the
low surface brightness and the contribution of small, transiently heated
grains. If the surface brightness is low, the noise
and/or the systematic errors produced in the subtraction of zodiacal
emission may make it impossible to derive an accurate optical depth. It
would be worthwhile to check the applicability of the method at high galactic
latitudes. The second problem may be resolved by using far-IR data at
wavelengths longer than 100~\mum.

\acknowledgements

We wish to thank Dr.~Y. Fukui, Dr.~A. Mizuno,
and the graduate students of the Department of
Physics and Astrophysics, Nagoya University for their help with
the CO line observations.
We are very grateful to Dr.~W. T. Reach and the anonymous referee for
their comments and suggestions.
We also thank Dr.~Hwankyung Sung for helpful discussions, 
and Mr.~Hyo-Ryung Kim for providing some useful CO line spectra
for this work.
This work has been supported in part by the 1994
% Korea Science and Engineering Foundation (KOSEF)
KOSEF International Cooperative Research Fund.

%\clearpage
\vskip 1.5cm

% REFERENCES

\clearpage
 
%Tables
 
\begin{deluxetable}{ccccc}
%\footnotesize
% \tablenum{1}
% \tablecolumns{5}
\tablecaption{POSITIONS OF ONE-DIMENSIONAL CUTS
AND OFF POSITIONS \label{tbl-1}}
\tablewidth{0pt}
\tablehead{
  &   & \multicolumn{2}{c}{$l$-range}  &
\colhead{OFFs\tablenotemark{a}} \\
\cline{3-4}
\colhead{Scan Number} & \colhead{$b$} &
\colhead{HI 21 cm line} & \colhead{CO J=1$-$0 line} & \colhead{($l, b$)}
}
\startdata
B1 & 1.\degrs0 & 42.\degrs5$-$48.\degrs0 & 43.\degrs0$-$45.\degrs5
& (45.\degrs0, 1.\degrs5), (43.\degrs5, 2.\degrs5) \nl
B2 & 2.0 & 43.0$-$45.7 & 43.0$-$46.0
& (45.0, 1.5), (43.5, 2.5) \nl
B3 & 3.0 & 42.5$-$48.0 & 44.0$-$47.0
& (43.5, 2.5), (45.8, 4.2) \nl
B4 & 4.0 & 43.0$-$47.5 & 44.0$-$46.5
& (47.5, 3.5), (45.8, 4.2) \nl
B5 & 5.0 & 42.5$-$50.0 & 45.0$-$48.0
& (46.5, 5.3), (49.0, 5.5) \nl
B6 & 6.0 & 42.5$-$51.5 & 46.0$-$48.5
& (46.5, 5.3), (49.0, 5.5) \nl
B7 & 7.0 & 42.5$-$51.5 & 46.5$-$47.5
& (45.5, 6.5) \nl
B8 & 8.0 & 42.5$-$51.0 & 46.5$-$47.5
& (48.0, 8.0) \nl
B9 & 9.0 & 42.5$-$51.5 & \nodata & \nodata \nl
\enddata
\tablenotetext{a}{OFF positions for CO J=1$-$0 line observations}
\end{deluxetable}

\begin{deluxetable}{cccccc}
%\footnotesize
% \tablenum{2}
\tablecaption{REGIONS WITH CO J=1$-$0 LINE EMISSION \label{tbl-2}}
\tablewidth{0pt}
\tablehead{
   &  &  & \colhead{\vlsr\tablenotemark{a}} &
\colhead{\tr\tablenotemark{a}} &
\colhead{\delvfwhm\tablenotemark{a}} \\
\colhead{Region} & \colhead{$b$} & \colhead{$l$-range} &
\colhead{(km~s$^{-1}$)} & \colhead{(K)} & \colhead{(km~s$^{-1}$)}
}
\startdata
A & 1.\degrs0 & 43.\degrs50$-$43.\degrs70 &
+60 & 2.5 & 2.1 \nl
B & 1.0 & 44.50$-$44.75 & $-$10 & 2.1 & 2.9 \nl
C & 3.0 & 45.30$-$46.65 & +27 & 3.0 & 7.8 \nl
D & 4.0 & 44.20$-$45.50 & +22 & 3.3 & 5.6 \nl
E & 5.0 & 47.65$-$47.95 & +14 & 1.8 & 2.5 \nl
\enddata
\tablenotetext{a}{Line parameters at peak position}
\end{deluxetable}

\clearpage
\onecolumn

% Figure Legends

% Fig. 1
\begin{figure}
\figurenum{1}
\epsscale{0.8}
\plotone{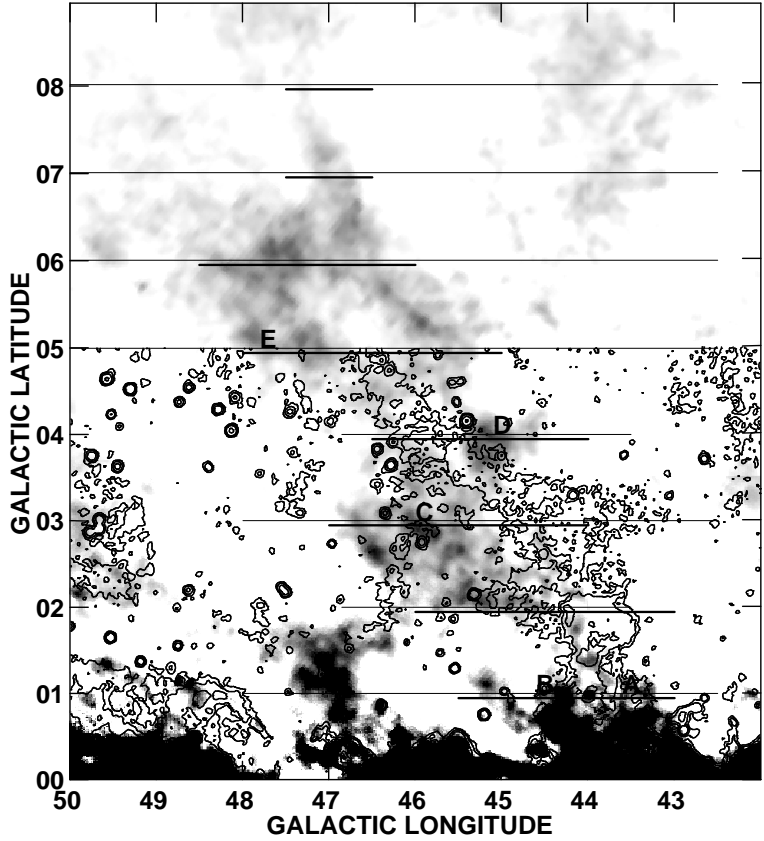}
%\plotfiddle{figure1.ps}{150}{90}{HSF=0}{VSF=100}{HTRANS=100}{VTRANS=0}
\caption{
100 \mum\ (grey scale) and 11~cm continuum (contour) images of \gw. 
The purpose of this figure is to show the morphology of \gw\ and to show
how the observed regions are related to it. We, therefore, have removed
a smooth background emission from the 100~\mum\ and radio continuum maps
using a median filter of 6\degr$\times$2\degr.
\gw\ is the far-IR structure extends straight up to $b$=7\degr\ from
the Galactic plane. The regions where HI and CO line data were obtained
are shown as light and dark solid lines, respectively.
For clarity, the dark solid lines are shifted in $b$-direction
from the light ones.
The regions with CO line emission are also labeled as A$-$E.
Grey scale intensity range is 1$-$30~MJy~sr$^{-1}$. 
Contour levels are 0.03, 0.1, 0.25, 1.0, and 3~K in brightness
temperature.}
\end{figure}

\clearpage

% Fig. 2
\begin{figure}
\vskip -3cm
\figurenum{2}
\epsscale{0.8}
\plotone{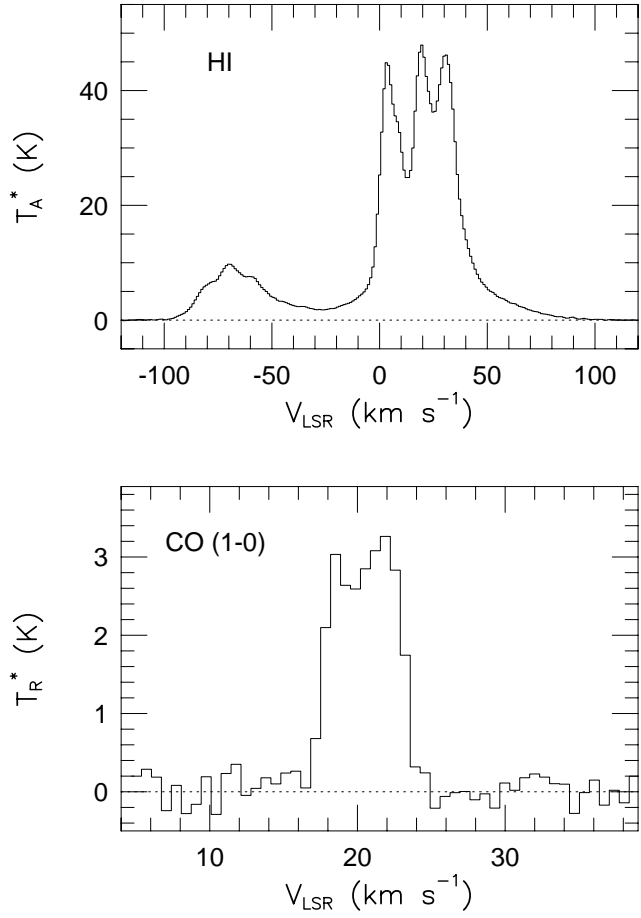}
\caption{
HI 21~cm and CO J=1$-$0 line profiles
at ($l$, $b$)=(45.\degrs10, 4.\degrs00) where the strongest CO line
emission
was detected. HI and CO lines were obtained using the Arecibo and
Nagoya 4~m telescopes, respectively.
The peaks between \vlsr$\simeq$18 and 40~\kms\ in the HI line profile
are the components associated with \gw.
The peak at \vlsr~$\simeq$~20~\kms\ has associated
molecular gas.}
\end{figure}

\clearpage

% Fig. 3
\begin{figure}
\figurenum{3}
\epsscale{1.0}
\plottwo{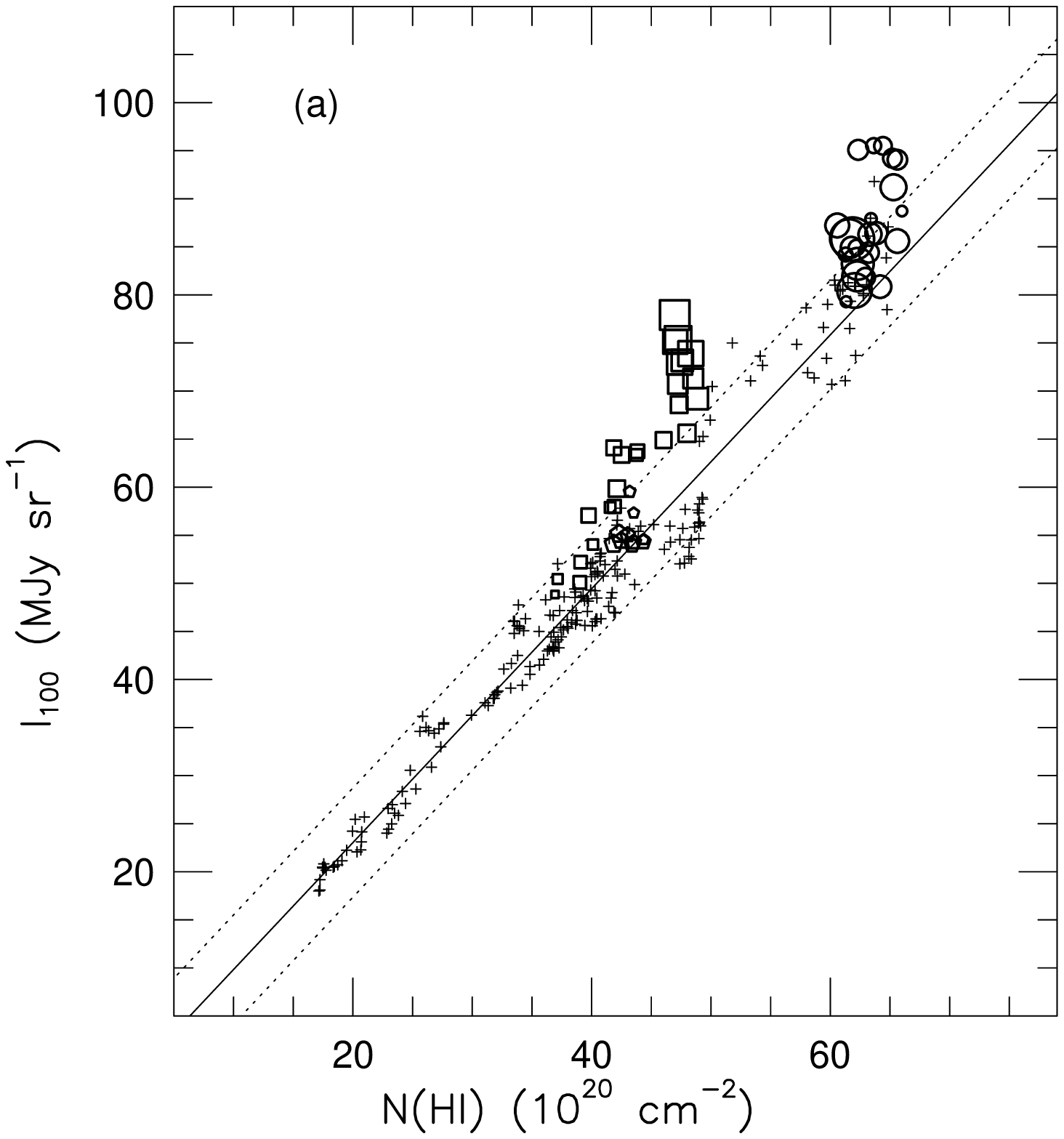}{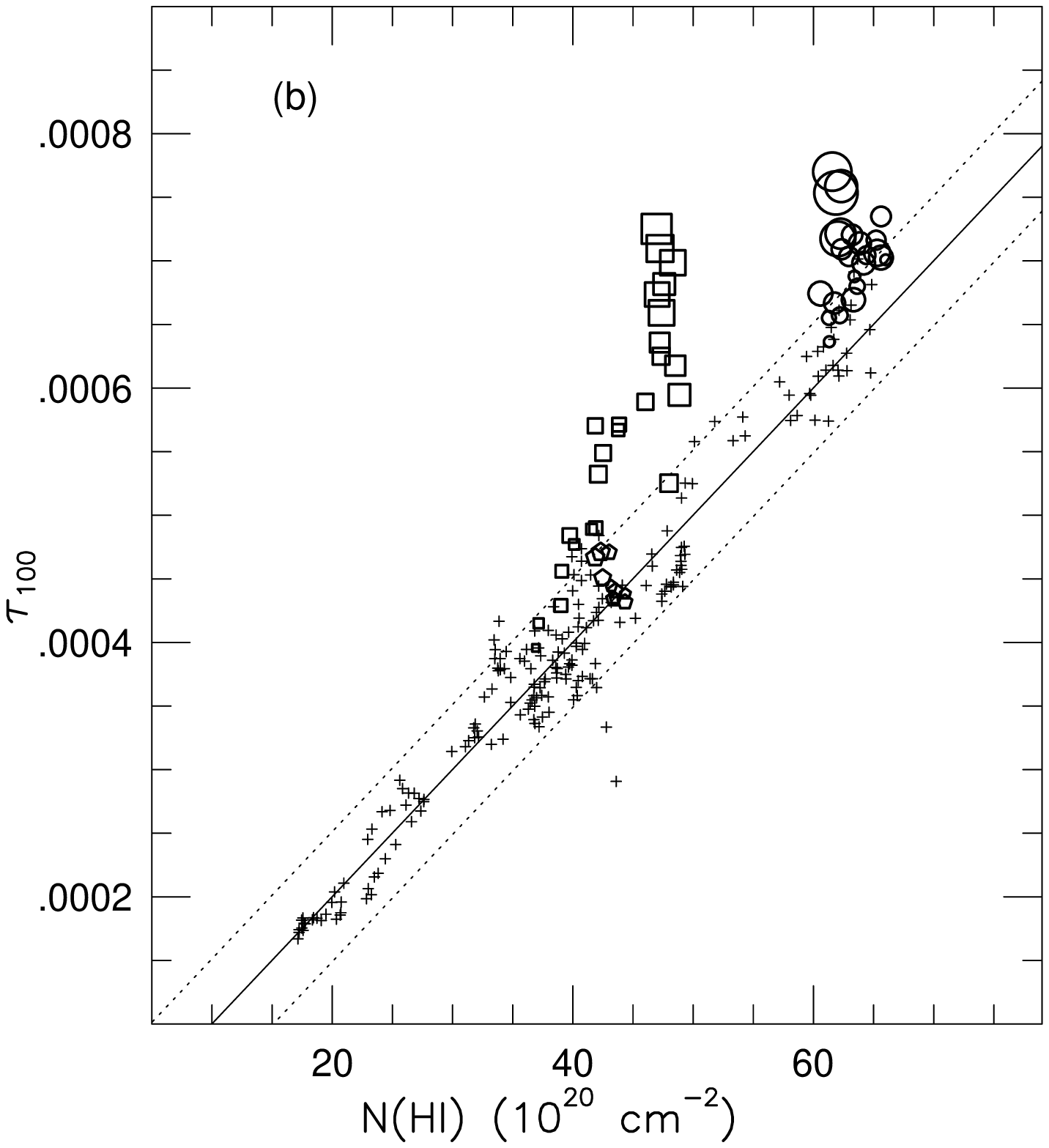}
\caption{
(a) \ihund\ vs. \nhi\ and (b) \thund\ vs. \nhi. 
%A good correlation is shown between the two parameters. 
Open circles, squares, and pentagons
represent points with detectable CO at $b$=3\degr, 4\degr, and 5\degr,
respectively. The area of each symbol is proportional to \wco.
Least-squares fits yield
(\ihund/MJy~sr$^{-1}$) = 1.32(\nhi/10$^{20}$~cm$^{-2}$) $-$ 3.37 and
(\thund/10$^{-5}$) = 1.00(\nhi/10$^{20}$~cm$^{-2}$) + 0.01, respectively. 
Results of the fits are shown as solid lines.
The dotted lines indicate $\pm$2$\sigma$
deviations from the least-squares fit. Note that the points with CO are
more clearly separated from the points without CO in (b) than in (a).
Also note that the areas of symbols become larger as the points become
further away from the fit in (b), whereas they do not in (a).}
\end{figure}

\clearpage

% Fig. 4
\begin{figure}
\figurenum{4}
\epsscale{1.0}
\plottwo{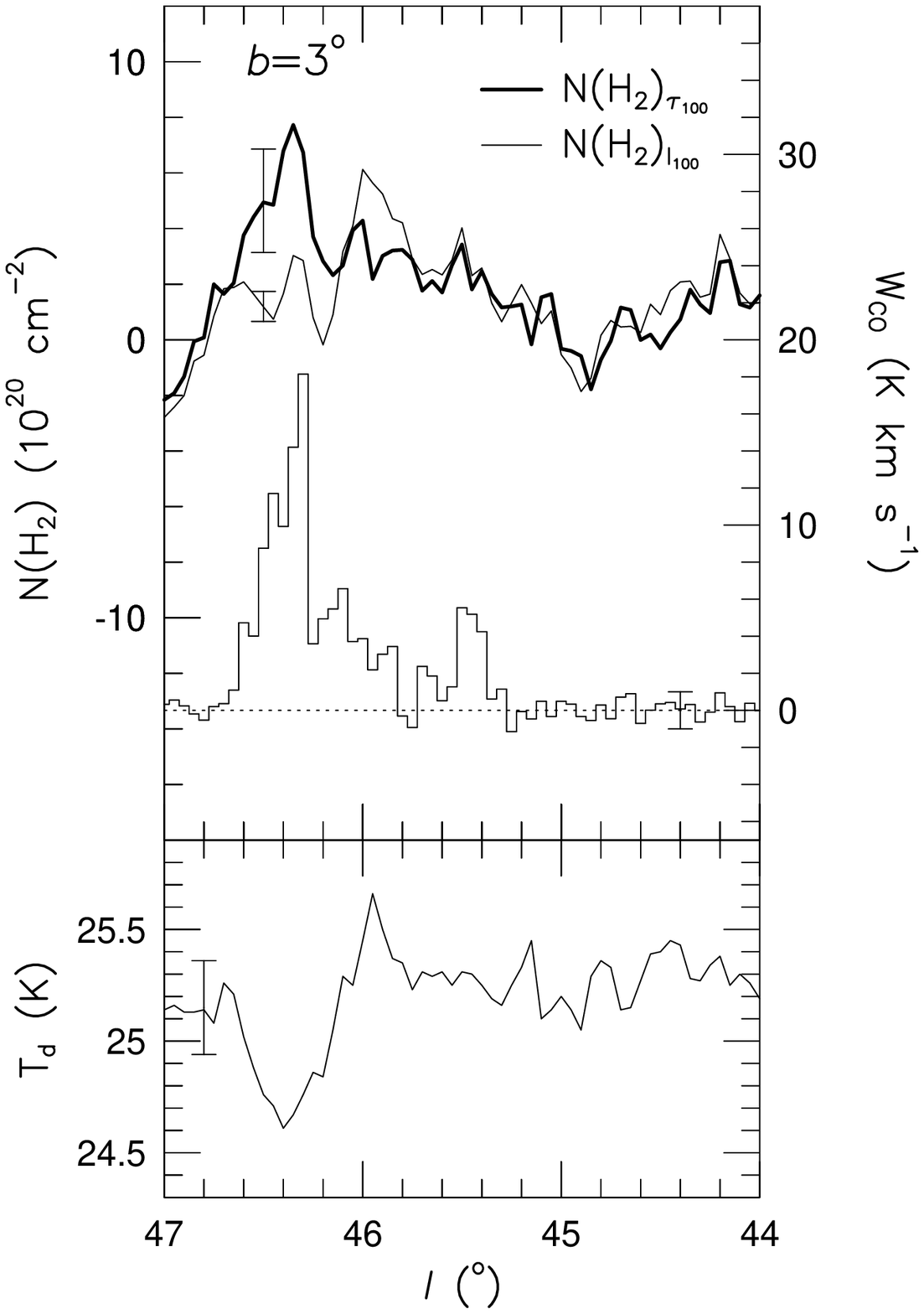}{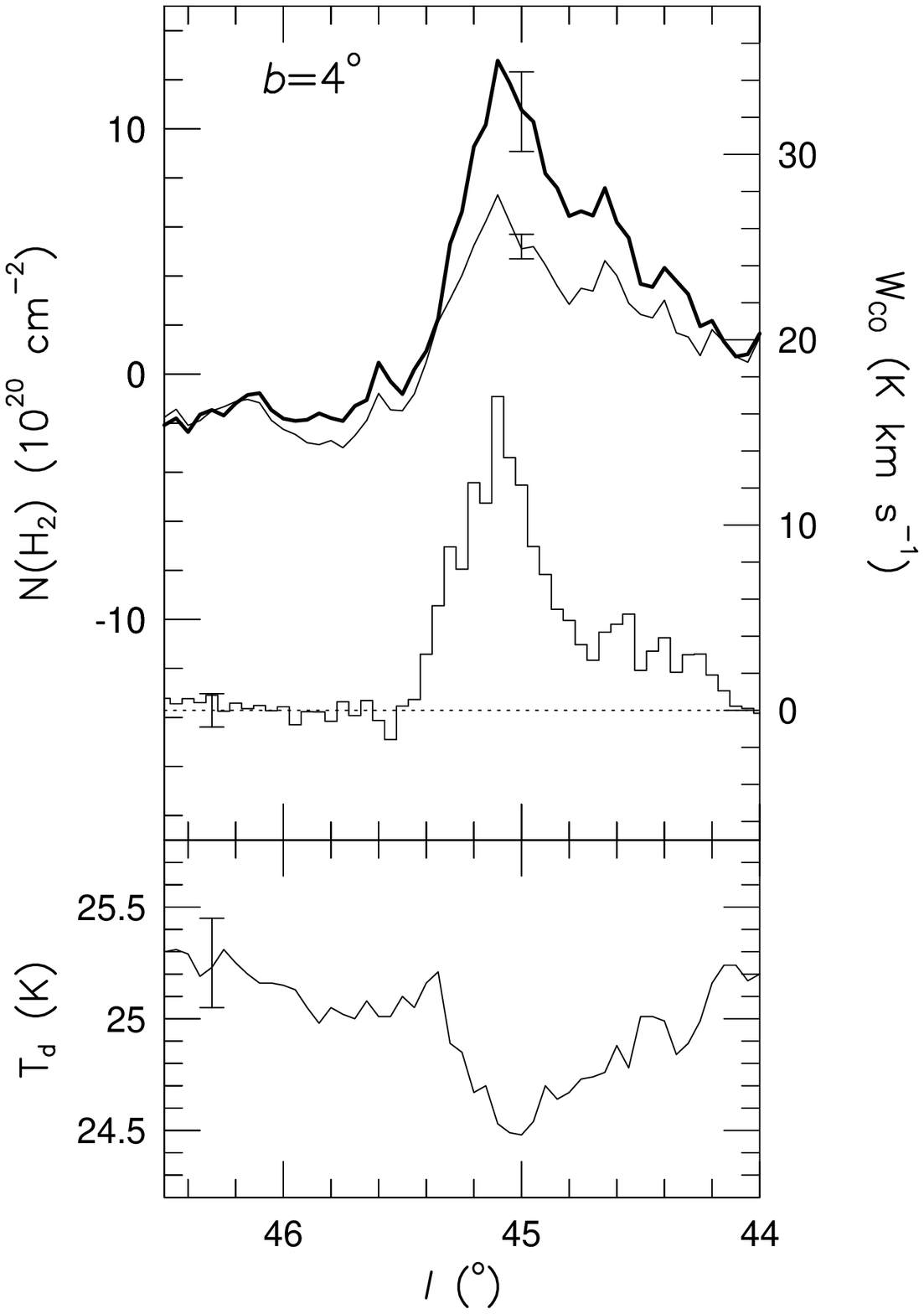}
\caption{
Distributions of \nhtwo$_{I_{100}}$, 
\nhtwo$_{\tau_{100}}$, \wco, and  $T_{\rm d}$ at (a) $b$=3\degr\ and 
(b) $b$=4\degr. Note that \nhtwo$_{I_{100}}$ differs considerably 
from either \wco\ or \nhtwo$_{\tau_{100}}$ at $b$=3\degr.
The difference is partly due to a local heating source.
It is also noticeable that dust temperature drops in the regions
where CO line emission is detected. 
The typical statistical error bars are given in each panel.}
% Each error bar represents typical statistical error.}
\end{figure}

\clearpage

% Fig. 5
\begin{figure}
\vskip -3cm
\figurenum{5}
\epsscale{0.8}
\plotone{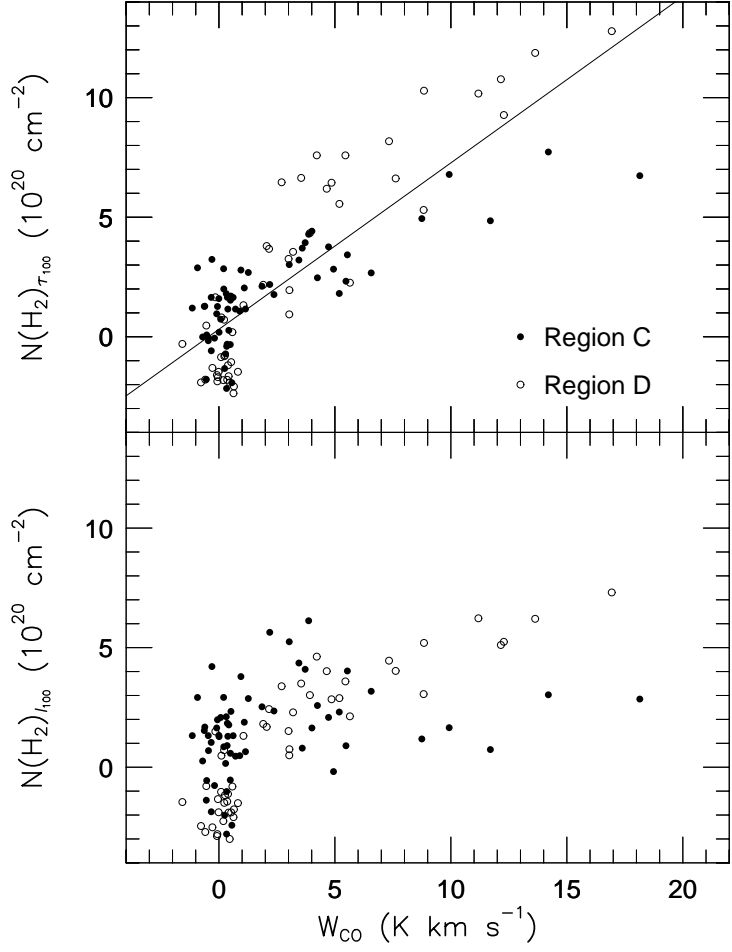}
\caption{
Comparison of \nhtwo$_{\tau_{100}}$ (upper panel) and
\nhtwo$_{I_{100}}$ (lower panel) with \wco.
Filled and open circles represent points at $b$=3\degr\ and 4\degr,
respectively. \nhtwo$_{\tau_{100}}$, but not
\nhtwo$_{I_{100}}$, has a very
good correlation with \wco. The straight line is a least-squares fit
to the data, 
(\nhtwo$_{\tau_{100}}$/10$^{20}$~cm$^{-2}$)
=0.70(\wco/K~\kms)$-$0.31.}
\end{figure}

\clearpage

% Fig. 6
\begin{figure}
\vskip -3cm
\figurenum{6}
\epsscale{0.8}
\plotone{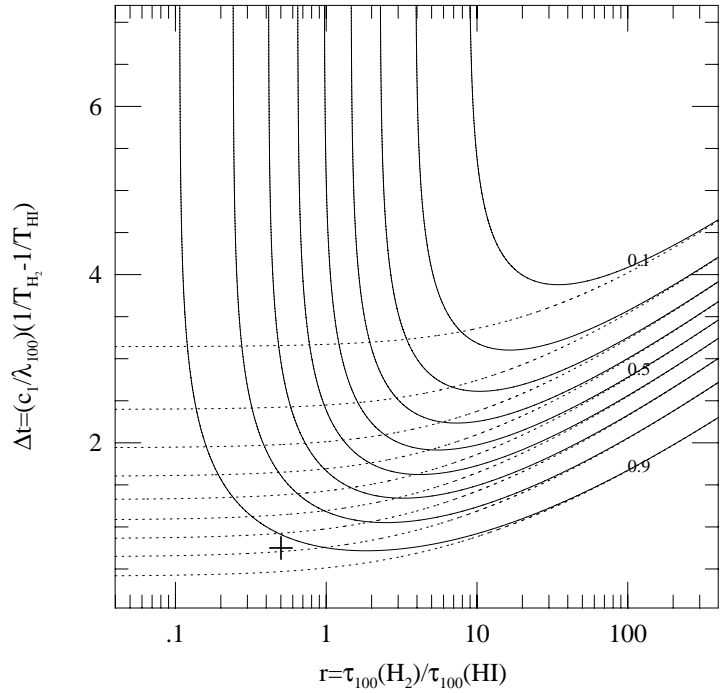}
\caption{
Variations of $\tau_{100}' / \tau_{100}$ (solid line) and
\nhtwo$_{\tau_{100}'}$/\nhtwo$_{\tau_{100}}$ (dotted line) in 
% ($r$, $\triangle t$)=$\Bigl( \tau_{100}({\rm H_2})/\tau_{100}({\rm HI}),
% ~(c_1/\lambda_{\rm 100})(1/{T_{\rm H_2}}-1/{T_{\rm HI}}) \Bigr)$ plane, 
($r$, $\triangle t$) plane. 
$r \equiv \tau_{100}({\rm H_2})/\tau_{100}({\rm HI})$
and $\triangle t \equiv$
($c_1/\lambda_{\rm 100}$)(1/{$T_{\rm H_2}$}-1/{$T_{\rm HI}$}), 
where $c_1/\lambda_{\rm 100}$=144.
The $\tau_{100}' / \tau_{100}$ ratio is weighted toward the warmer dust
associated with HI gas, 
but \nhtwo$_{\tau_{100}'}$/\nhtwo$_{\tau_{100}}$ is not.
The difference between the two quantities is larger at smaller $r$. 
The cross indicates the position of region D in ($r$, $\triangle t$) plane.
In order to estimate an upper limit to the error in \nhtwo$_{\tau_{100}}$
in region D, we assume (\thi, \thtwo)=(25.2~K, 22.2~K).
Contour levels are 0.1, 0.2, $\cdots$, 0.9.}
\end{figure}

\clearpage

% Fig. 7
\begin{figure}
\vskip -1cm
\figurenum{7}
\epsscale{0.8}
\plotone{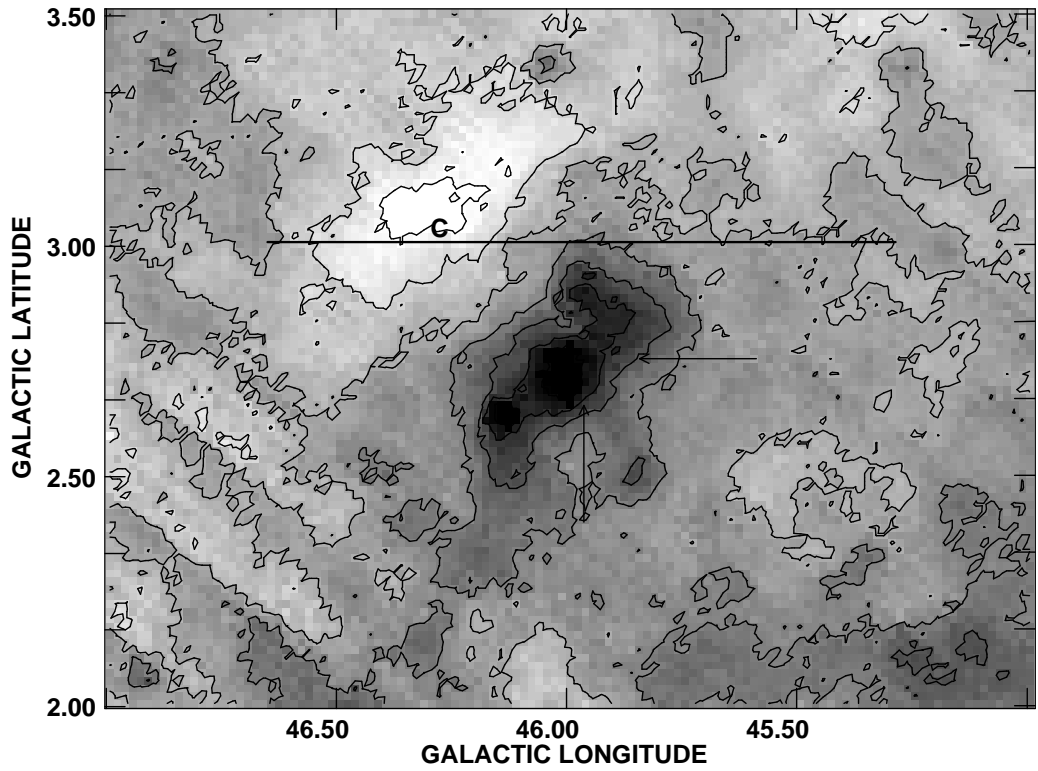}
\caption{
Distribution of \isixty/\ihund\ around region C.
A luminous blue giant star (LSII+12.\degrs3) is located where two
arrows meet. Note the high intensity ratio around the star.
The region where CO line emission is detected 
is shown as a solid line. Grey scale intensity range is 0.20$-$0.26.
The lowest contour level is 0.20 and contour spacing is 0.01.}
\end{figure}

\end{document}